\documentclass[12pt,a4paper,final]{iopart}
\usepackage{iopams}  
\usepackage{graphicx}
\usepackage[breaklinks=true,colorlinks=true,linkcolor=blue,urlcolor=blue,citecolor=blue]{hyperref}
	
\usepackage[symbol]{footmisc}

\usepackage[
backend=bibtex,
style=numeric-comp,
sorting=none,
block=none,
indexing=false,
citereset=none,
isbn=false,
url=true,
doi=true,
natbib=true,
]{biblatex}
\begin{document}

\title[Open-source neuronavigation for brain stimulation]{Open-source neuronavigation for multimodal non-invasive brain stimulation using 3D Slicer}

\author{Frank Preiswerk$^{1}$}
\address{$^1$Department of Radiology, Brigham \& Women’s Hospital, Harvard Medical School, USA.}
\ead{frank.preiswerk@gmail.com}

\author[cor1]{Spencer T. Brinker$^{1}$}
\address{$^1$Department of Radiology, Brigham \& Women’s Hospital, Harvard Medical School, USA.}
\ead{spencer.t.brinker@gmail.com}

\author{Nathan J. McDannold$^{1}$}
\address{$^1$Department of Radiology, Brigham \& Women’s Hospital, Harvard Medical School, USA.}
\ead{njm@bwh.harvard.edu}

\author{Timothy Y. Mariano $^{2,3,4}$}
\address{$^2$Butler Hospital, Providence, USA}
\address{$^3$Center for Neurorestoration and Neurotechnology, Providence Veterans Affairs Medical Center, Providence, USA.}
\address{$^4$Department of Psychiatry, Brigham \& Women’s Hospital, Harvard Medical School, USA.}
\ead{mariano@post.harvard.edu}

\begin{abstract}
In recent years, non-invasive neuro-modulation methods such as Focused Ultrasound (FUS) have gained popularity. The aim of this work is to introduce the use of existing open-source technology for surgical navigation to the field of multimodal non-invasive brain stimulation. Unlike homegrown and commercial systems, the use of well-documented, well maintained, and freely available open-source components minimizes the learning curve, maximizes technology transfer outcome, and fosters reproducible science for complex, guided neuromodulation systems. The described system significantly lowers the entry bar to clinical research and experimentation in the field of non-invasive brain stimulation. Our contribution is two-fold. First, a high-level overview of the components of the descried system is given in this manuscript. Second, all files are made available online, with a comprehensive step-by-step manual, quickly allowing researchers to build a custom system. A spatial accuracy of 0.93 mm was found through validation using a robotic positioning system.

\end{abstract}

\vspace{2pc}

\section{Introduction}

Non-invasive brain stimulation (NIBS) in humans via Neuronavigation-Guided (NG) techniques have recently gained significant traction for use in clinical practice and for research investigations. Preprocedural Magnetic Resonance Imaging (MRI) in combination with real-time optical tracking are used to guide energy from a brain stimulation transducer device to patient-specific brain MRI targets. A crosshair type virtual point is assigned to the Spatial Peak Intensity Location (SPIL) in the critical area of the transmitting energy being emitted from the transducer. The virtual point is then overlaid onto the MRI imaging volume for real-time visualization of the SPIL position in the brain during transducer operation. Among the most popular modalities are Transcranial Magnetic Stimulation (TMS) and transracial Focused Ultrasound (tFUS). TMS is now used clinically as a treatment option in psychiatry and is FDA approved to treat major depressive disorder \cite{carpenter2012transcranial} and obsessive-compulsive disorder \cite{FDA2008}. Multiple vendors offer NG for use in TMS platforms like Brainsight TMS Navigation (Rogue Resolutions Ltd, United Kingdom) and visor2™QT (ANT Neuro, Netherlands). Academic laboratory investigators have also developed custom NG for TMS \cite{souza2018development}. Low-intensity tFUS is an up and coming method for experimental human neuromodulation \cite{fomenko2018low}. These custom NG tFUS systems have been used to target locations within the human cortex \cite{lee2016simultaneous} and thalamus \cite{legon2018neuromodulation}. At present, no commercial NG tFUS commercial system is available on the market.

Planning and navigating neuromodulation procedures requires complex real-time software to track the relative position of multiple tracking objects attached to the transducer and the patient for integrating these locations onto the 3D space containing the patient’s MRI volume. There are commercial surgical navigation platforms available used for intra-operative image fusion including systems by BrainLab Inc, (Munich, Germany) and Stryker Inc, (Kalamazoo, MI, USA). Aside from their high cost, such systems are often tailored to and approved for specific interventions. These proprietary systems can be difficult to integrate into a research project and therefore are often ill-suited for research and experimentation of new applications. Some researchers have developed custom hardware and software systems from scratch in the laboratory \cite{kim2012image}. However, as recently noted in an article regarding current technical challenges for brain ultrasound applications, “homegrown” systems have a reproducibility problem \cite{landhuis2017ultrasound}. They can be hard to operate, maintain, and reestablish after the tenure of their creator due to lack of internal laboratory technology transfer or limited published documentation of system design. Our system aims to fill this technology gap by providing non-invasive neuromodulation built on top of a solid foundation consisting of well-documented and maintained open-source software packages. 

Existing software and modules originally intended for surgical navigation, 3D visualization, image registration, and real-time data processing are configured specifically for neuromodulation applications where a subject, SPIL and a stylus need to be tracked in real-time. We have designed our open-source system so a pre-configured 3D Slicer workflow can be set up with minimal knowledge of NG systems and little effort, yet all components are highly customizable. The proposed system covers general use cases for neuronavigation. Head movements of the subject are tracked in real-time while the SPIL of an arbitrary modality (e.g., tFUS) is rendered onto pre-obtained MRI. Co-registration of the subject and their brain MRI is conducted via fiducial markers placed on the subject’s head. Re-registration on a subsequent day does not require the subject to obtain another MRI. The spatial accuracy of the system is validated using a robotic positioning system. A detailed manual is freely provided online, which serves as a starting point for adapting the system to custom experiments. The proposed platform is a first push towards reproducible science in this emerging field, and it will further enable shared data, methods, and results across the community.

\section{Material and Methods}

\subsection{System overview}

The minimum components required to implement the methods described in this manuscript are as follows: A computer (64-bit Windows, at least 4GB RAM), a position tracking device supported by Plus toolkit (NDI Polaris Vicra optical tracking camera is used here), one tracker for head, applicator (e.g., FUS), and pointer device (stylus), respectively. Furthermore, attachable fiducial markers for the pre-operative image (such as Gadolinium-based markers for MRI) are required.

The described system was built on a computer workstation (Intel Core i5-6500, 3.2GHz, 8GB Ram, 64-bit Windows 7) connected to an NDI Vicra optical tracking camera (NDI Medical, Cleveland, Ohio, USA). 3D Slicer (version 4.8.0) \cite{fedorov20123d} and the Plus Toolkit (version 2.6.0) \cite{lasso2014plus} form the core of the navigation software. 3D Slicer serves as the user interface as well as the central software framework for data and image processing and real-time IO using the OpenIGTLink network protocol \cite{tokuda2009openigtlink}, an open standard for network communication among medical devices. Within 3D Slicer, the SlicerIGT extension is employed, which contains additional functionality for rapid prototyping of applications for image-guided interventions. Plus Toolkit and 3D Slicer communicate tracking data over a network connection using the OpenIGTLink protocol. Plus Toolkit supports a range of different tracking systems including electromagnetic and optical systems. Although an optical tracking camera is used here, the abstraction provided by Plus toolkit allows one to substitute any other supported tracking device with minimal effort. All components are actively maintained open-source packages under BSD-style licenses, which allows great flexibility not only for academic but also for commercial use.

\subsection{3D Slicer}

3D Slicer is an open source software platform for medical image informatics, image processing, and three-dimensional visualization. It has built-in support for import / export of a wide range of 2d and 3d medical image formats. It can render 2d and 3d images alongside common graphics primitives such as points (landmarks) and polygon models, e.g., neuromodulation applicators and SPIL foci. In 3D Slicer, all objects are represented as nodes in a scene graph, an established concept from computer graphics for representing objects and transformations. Nodes can be nested arbitrarily, and all transform nodes in a branch are applied to all children. The state of the scene graph controls the 2D and 3D rendering windows. 3D Slicer’s built-in OpenIGTLink networking functionality allows interfacing with other applications including Plus Toolkit, as described below. When used as a surgical navigation system, tracking information for each tracked tool received via the network is used to continuously update the corresponding transforms in the scene graph. For the proposed system, the SlicerIGT extension for 3D Slicer \cite{ungi2016open} was installed via 3D Slicer’s Extension Manager. It comprises a set of tools for rapid prototyping of IGT solutions within 3D Slicer. Fig. 1 shows an overview of the hardware and software components of the neuro-modulation navigation system proposed here.

\begin{figure}[htb!]
 \centering
 \includegraphics{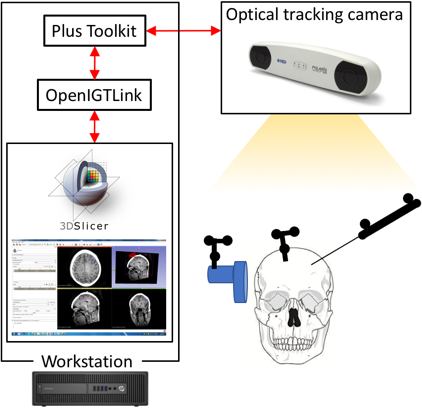}
 \caption{\label{figureone}Overview of the navigation system for non-invasive neuro-modulation. A camera is tracking the 6 degrees of freedom of three optical markers: A marker attached to the patient, pointer tool (stylus), and the applicator (e.g., a focused ultrasound transducer or transcranial magnetic stimulation coil). Each marker consists of a set of 3-4 optically reflective spheres. }
\end{figure}

\subsection{Plus Toolkit}

Plus Toolkit is an open-source software package designed to simplify the acquisition and synchronization of various signal sources in image-guided therapy, including real-time imaging and tracking information. It provides an abstraction layer for low-level data acquisition by connecting to many different vendor APIs, and forwarding the data in a standardized format to other applications over a network connection [9]. For the purpose of the described navigation system, Plus Toolkit is used exclusively to obtain tracking information from the camera in real-time, as depicted in Fig. 1. For each tracked object, a 4x4 homogenous transformation matrix that encodes all six degrees of freedom for a tracked object is provided from Plus Toolkit via OpenIGTLink in real-time (10 fps here). Plus Toolkit runs as a standalone application and interacts with 3D slicer through a TCP/IP connection. It is configured using an XML file which, for our purpose, consists of two main parts. First, the tracking device and all tracked tools (by reference to ‘.rom’ files) are specified. Second, it contains the configuration for the OpenIGTLink server, spanned by Plus Toolkit, that sends the tracking data out via TCP/IP. The “Plus Server” application has to be started first, and the configuration file has to be loaded initially. It then runs in the background and provides tracking data to any application that connects to its OpenIGTLink server. Once 3D Slicer is started, its built-in OpenIGTLink interface module is used to connect to the Plus Toolkit’s OpenIGTLink server (Fig. 1).

\subsection{Custom tracked tools}

For surgical guidance, all relevant tools must be tracking-enabled. In the case of optical tracking used here, objects are equipped with a set of passive optically reflective spheres. In order to distinguish different objects, each set of spheres must be attached in a geometrically different way. Attachable markers and a stylus can be obtained directly from the manufacturer of the tracking camera or they can be custom made with little effort. In our system, an attachable tracking marker for an applicator as well as a complete tracked pointing tool were custom made from resin using a laser cutter, as depicted in Fig. 2. Optical spheres and their screw bases were obtained from the manufacturer (NDI passive spheres, Northern Digital, Inc, Ontario, Canada) and screwed into the resin. The geometry of each tool was obtained using the proprietary software “6D Architect”, provided by the manufacturer (similar software is available from other tracking camera manufacturers). The output of this process is a ‘.rom’ file for each tool, which is later required for and referenced in the configuration file for Plus Toolkit (as described above).

\subsection{Point-based registration and landmarking using SlicerIGT 3D Slicer Module}

Point-based registration is a method to register a tracked object to its virtual counterpart in 3D Slicer. Among many other functions, the SlicerIGT Module provides a convenient user-interface for obtaining corresponding landmarks on image data via Slicer’s 3D view, as well as from the physical world using the stylus. From this set of corresponding landmarks, SlicerIGT computes the transformation matrix from the reference to 3D Slicer’s RAS (right/anterior/superior) coordinate system, ReferenceToRAS. It serves as the link between all tracked objects in physical space and 3D Slicer’s coordinate system and forms the root node of the scene graph. The landmarks in image data are typically obtained from fiducial markers attached to the subject during acquisition of the volumetric image, such as gadolinium markers in MRI.

Similar to the subject, the applicator is registered to its rigidly attached marker. A 3D model of the applicator is loaded into 3D Slicer in one of its supported file formats (e.g., ‘.obj’ or ‘.vtk’). Corresponding landmarks are selected on the applicator using the stylus, and on the 3D model in 3D Slicer, and the transform is generated from these correspondences. Lastly, the applicator’s focus is determined (using an application-dependent measuring device and the stylus), to obtain its location in the Applicator’s coordinate frame (instructions on how to find the spatial peak for a FUS pressure field are provided in the manual). Both applicator and focus calibrations only have to be performed once and can be saved as part of the 3D Slicer scene. Plus Toolkit provides the transforms of both applicator and stylus expressed in the Reference coordinate system, which are appended in the scene graph below ReferenceToRAS, as depicted in Figure 2.

\begin{figure}[htb!]
 \centering
 \includegraphics{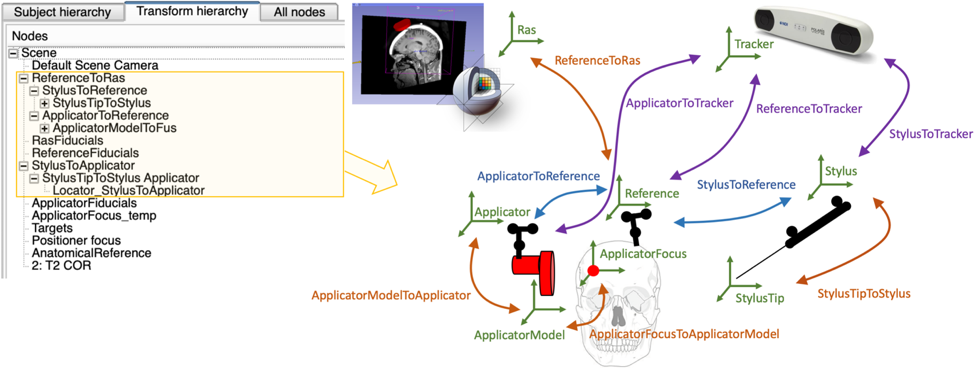}
 \caption{\label{figuretwo} Transform hierarchy inside 3D slicer (left), and illustration of the transforms (right). All tools are tracked in the tracker’s coordinate frame (purple transforms). Plus Toolkit is configured to send the transforms of all tracked tools via OpenIGTLink relative to the reference coordinate system (blue transforms). Orange transforms show fixed transforms that are the result of all registration procedures. In particular, ReferenceToRas defines the mapping from all physical objects to 3D Slicer’s RAS (right, anterior, superior) coordinate system.}
\end{figure}

\subsection{Pre-configured 3D Slicer scene and detailed manual}

While high-quality documentation for all described software components is available online, there is still a need, specifically for FUS researchers, to have a simpler way to tie all components together and have them up and running with little effort and in a short amount of time. As part of this technical note, a complete 3D Slicer scene with configuration files for Plus Toolkit is provided along with detailed step-by-step instructions on how to perform all fiducial registrations necessary to build a custom navigation system quickly, via GitHub at  \url{https://github.com/fpreiswerk/neuronav}.

\subsection{Experimental validation}

The isotropic spatial accuracy of the NDI Polaris Vicra system used here (NDI Polaris Vicra, Northern Digital, Inc, Ontario, Canada) is 0.35 mm RMS (0.5-mm 95\% confidence interval), as reported by the vendor’s performance specifications assessment. However, the vendor’s accuracy measurement should be regarded as a lower bound for the system proposed here, built with custom tools. Therefore, the navigation system was validated using a linear and motion-control positioning system (MB603601J-S6, Velmex Inc., Bloomfield, NY), which has an accuracy of 0.076 mm as reported by the manufacturer. All eight corner positions of a 20 x 20 x 20 mm cube were scanned by the positioning system, and simultaneously recorded with optical tracking camera using the stylus. One cube corner was used as the reference (p1), and the distance to the other 7 corners (p2-p8) was computed and compared to the distances of a perfect cube of the same size (20 x 20 x 20 mm).
Both reference marker and stylus used for this validation were custom made as described above. Fig. 3 depicts the validation setup. With these custom-made tools, the isotropic spatial accuracy was measured to be 0.93 mm (SD 0.74 mm) by comparative analysis of the eight measured points in 3-dimensional space.

\begin{figure}[htb!]
 \centering
 \includegraphics{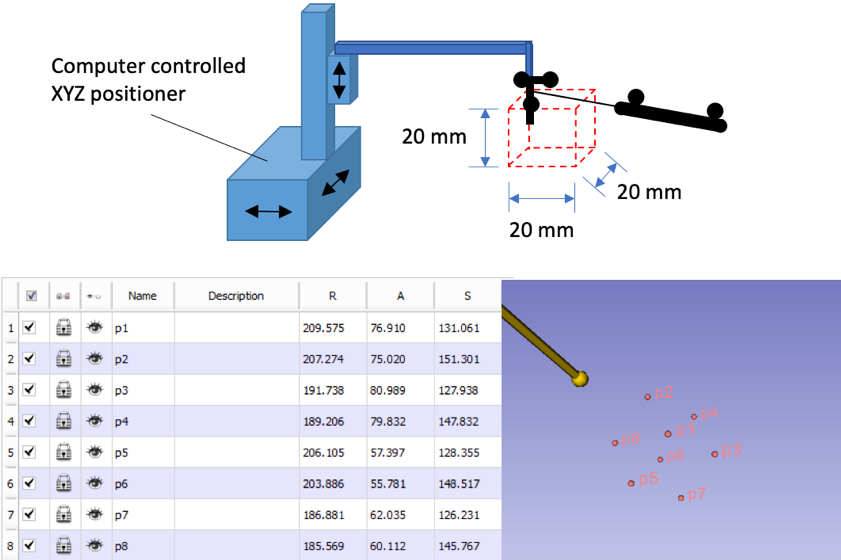}
 \caption{\label{figurethree} The setup used for validation. Top) A robotic system scanned the corners of a cube, and the positions were recorded using custom tracking tools. The tracking reference and the stylus were both custom-made. Bottom) View of the cube corners (coordinates left, 3d rendering right) and the real-time stylus position (yellow) in 3D Slicer.}
\end{figure}

\section{Conclusion}

A navigation system for non-invasive brain stimulation built using existing open-source components is proposed. It leverages the tremendous work from the image-guided therapy community, and it is meant as an initial contribution to guide the neuromodulation community towards a similar open approach to image-guidance and reproducible science. The system significantly reduces the time needed to get started with experimental prototypes for neuromodulation procedures. While this manuscript gives an overview on the components, an important part of this contribution is a set of scene and configuration files and a detailed manual available at \url{https://github.com/fpreiswerk/neuronav}. The scene and configuration files for 3D slicer and Plus Toolkit serve as a scaffolding to quickly build a custom system, and the detailed manual guides one through the process step by step. Application-specific markers can be built with little effort, and the accuracy of 0.93 mm found with the custom markers demonstrates that custom markers are accurate enough for typical neuromodulation procedures such as tFUS. 

\section*{Acknowledgements}
Financial support from NIH grants PO10CA174645, R25CA089017, R03EB025546 and a BBRF 2015 NARSAD Young Investigator Grant is duly acknowledged. The contents do not represent the views of the Department of Veterans Affairs or the United States Government. TYM is currently employed by Sage Therapeutics, Inc. and has previously consulted for Janssen Pharmaceuticals Inc and Ad Scientiam SAS; none of these supported or influenced the present work. The authors thank Jason Phillip White, PhD for assistance with the accuracy measurements, as well as the 3D Slicer and Plus toolkit community for developing and maintaining their highly valued open-source projects. 

\medskip

\printbibliography

\end{document}